\documentclass[runningheads]{llncs}
\usepackage{graphicx}
\usepackage{booktabs}
\usepackage{multirow}
\usepackage{subcaption}
\usepackage{adjustbox}
\usepackage{setspace}
\usepackage{amsmath}
\usepackage{xspace}
\usepackage{amssymb}
\usepackage{wasysym}
\usepackage{enumitem}
\usepackage{tabularx}
\usepackage{caption}
\usepackage{makecell}  

\usepackage{booktabs}  
\usepackage{threeparttable} 
\usepackage{hyperref}
\hypersetup{
    colorlinks=true, 
    linkcolor=blue,  
    filecolor=blue,  
    urlcolor=blue,   
    citecolor=blue,  
}
\usepackage{subcaption}
\begin{document}

\title{Spatial-aware Attention Generative Adversarial Network for Semi-supervised Anomaly Detection in Medical Image}
\titlerunning{SAGAN} 

\author{Zerui Zhang\inst{1}
Zhichao Sun\inst{1}
Zelong Liu\inst{1}
Bo Du\inst{1}
Rui Yu\inst{3}
Zhou Zhao\inst{2}
Yongchao Xu\inst{1}
}
\authorrunning{Zerui Zhang et al.}

\institute{$^{1}$ School of Computer Science, Wuhan University, Wuhan, China\\
\email{yongchao.xu@whu.edu.cn}\\
$^{2}$ School of Computer Science, Central China Normal University, Wuhan, China\\
\email{zhaozhou@ccnu.edu.cn}\\
$^{3}$ University of Louisville, Louisville, KY, USA\\
\email{rui.yu@louisville.edu}\\
}

\maketitle

\begin{abstract}
Medical anomaly detection is a critical research area aimed at recognizing abnormal images to aid in diagnosis. 
Most existing methods adopt synthetic anomalies and image restoration on normal samples to detect anomaly. The unlabeled data consisting of both normal and abnormal data is not well explored.
We introduce a novel \textbf{S}patial-aware \textbf{A}ttention \textbf{G}enerative \textbf{A}dversarial \textbf{N}etwork (SAGAN) for one-class semi-supervised generation of health images.
Our core insight is the utilization of position encoding and attention to accurately focus on restoring abnormal regions and preserving normal regions. To fully utilize the unlabelled data, SAGAN relaxes the cyclic consistency requirement of the existing unpaired image-to-image conversion methods, and generates high-quality health images corresponding to unlabeled data, guided by the reconstruction of normal images and restoration of pseudo-anomaly images.
Subsequently, the discrepancy between the generated healthy image and the original image is utilized as an anomaly score.
Extensive experiments on three medical datasets demonstrate that the proposed SAGAN outperforms the state-of-the-art methods. Code is available at
\href{https://github.com/zzr728/SAGAN}{https://github.com/zzr728/SAGAN}

\end{abstract}
\section{Introduction}

The early anomaly detection of medical images plays a crucial role in the diagnosis and treatment of various pathologies. To alleviate the physician's reading burden and enhance diagnostic efficiency, computer-aided diagnosis (CAD) systems employing deep learning are gaining popularity~\cite{luo2020deep}.
However, obtaining annotations is consistently challenging in real-world medical datasets due to the scarcity of anomalous data. This scarcity motivates the development of unsupervised anomaly detection (UAD), which operates without requiring annotations.
Current UAD methods view anomaly detection as a one-class classification (OCC) problem~\cite{OCC}. In classical UAD approaches, only normal images are used for training to model the distribution of normal images. During the testing phase, out-of-distribution samples are identified as anomalies.

UAD methods can be roughly categorized into reconstruction-based, embedding-based, restoration-based and self-supervised-based methods. Reconstruction-based methods learn the latent space representation of normal data, and identify abnormal data by their reconstruction errors. These methods use autoencoder~\cite{IGD,MemAE,SALAD,IF,MorphAEus} or Generative Adversarial Networks (GANs) ~\cite{AnoGAN,f-anogan,Ganomaly}. Embedding-based methods~\cite{MKD,panda,swssl} primarily leverage discriminative features to model normal sample distributions in the feature space. Self-supervised-based methods propose synthetic anomalies to model normal distribution and detect abnormal structure~\cite{pii,fpi,cutpaste,schluter2022natural}. Restoration-based methods~\cite{diffusion,healthygan,reversing,squid,tang2021disentangled} aim to restore pathological regions to healthy structures.
However, the capacity to detect medical anomalies is significantly hindered by the absence of a large number of realistic anomaly images. Furthermore, a substantial amount of unlabeled data containing genuine anomalies exists within real-world medical datasets, yet these  unlabeled images are overlooked by existing methods.


Recently, several approaches have highlighted the issue of underutilized unlabeled data~\cite{amae,DDAD,healthygan,brainomaly}. The main concept behind Dual-distribution Discrepancy for Anomaly Detection (DDAD)~\cite{DDAD} and Adaptation of pre-trained Masked AutoEncoder (AMAE)~\cite{amae} involves training dual-distribution reconstruction networks to gauge the differences in reconstruction between abnormal or unlabeled data and normal data. However, these approaches typically necessitate two phases and multiple networks, making it more cumbersome. HealthyGAN~\cite{healthygan} and Brainomaly~\cite{brainomaly} utilize normal images as supervision for generating healthy images with their unlabeled counterparts. Nevertheless, relying solely on normal image supervision is inadequate, as it may result in generating a healthy image that fails to restore abnormal regions on the original image, thus increasing the occurrence of false positives. Additionally, these methods overlook the recurrent anatomical structures present in most radiography images~\cite{squid}. 
 
In this work,
we propose a \textbf{S}patial-aware \textbf{a}ttention \textbf{G}enerative \textbf{A}dversarial \textbf{N}etwork (SAGAN) method for semi-supervised anomaly detection in medical images. 
SAGAN supervises the restoration of unlabeled data by ensuring accurate reconstruction of normal data and precise restoration of pseudo-anomaly regions. Recognizing the anatomical consistency inherent in most medical images, SAGAN divides the input image into \textit{N} patches and incorporates positional conditional encoding for each patch. This strategy leverages the similarities found in structures at the same position, thereby enhancing the restoration quality.

\begin{figure}[h]
    \centering
    \includegraphics[width=1.0\textwidth]{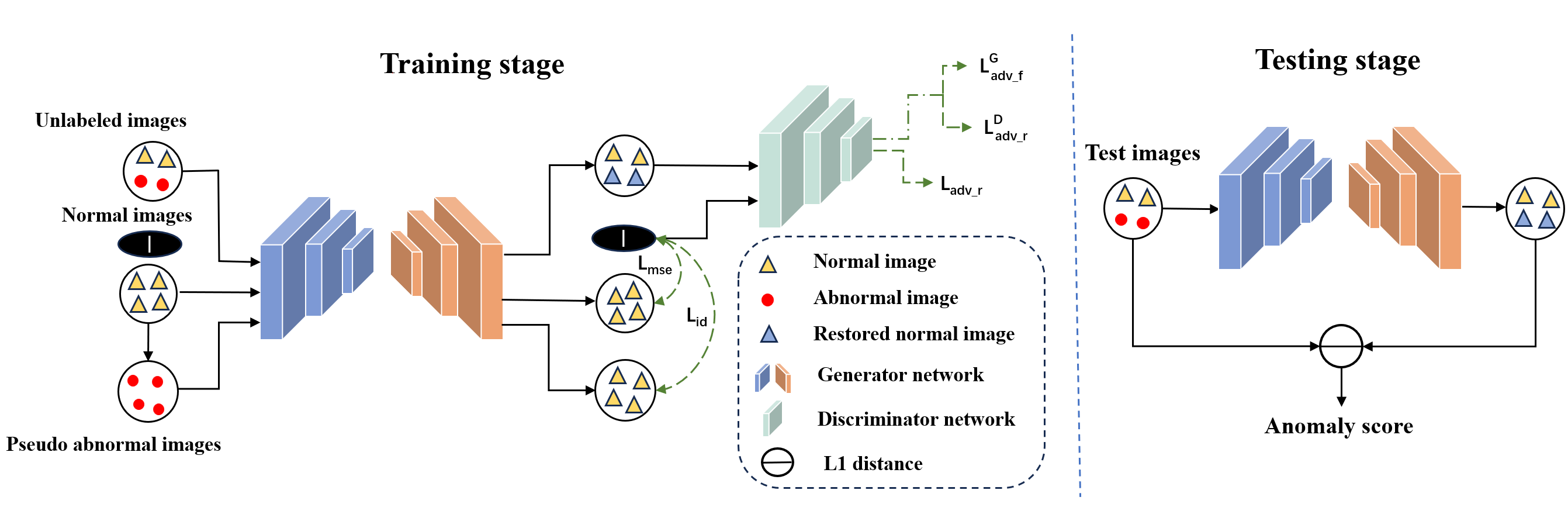}
    \caption{Overview of SAGAN. During the training stage, SAGAN learns to retore unlabeled images as normal images with the supervision of reconstructing normal images and restoring pseudo abnormal images. At the testing stage, the difference between the generated image and the original image reveals the presence of anomaly data.}
    \label{fig:whole_pipeline}
\end{figure} 

Moreover, the primary challenge in unlabeled image restoration lies in accurately restoring the anomalous region rather than merely generating a healthy image, which might possess a structure entirely different from the original one.
To address this issue, SAGAN incorporates an attention gate mechanism inspired by prior work~\cite{attentiongate}, which dynamically emphasizes regions of interest and generates corresponding masks. These masks enable the SAGAN decoder to retain normal areas while effectively discarding abnormal regions of interest and generating the corresponding normal structures. Through extensive experimentation, we achieve state-of-the-art performance on three challenging medical datasets.

\section{Methods}

\subsection{Overview}
In the propoesd SAGAN as shown in Fig.~\ref{fig:whole_pipeline}, we use a GAN-style framework to generate near-realistic healthy images from unlabeled images, healthy images and pseudo abnormal images. 
In concrete terms, the generator generates an additive map where each pixel represents the difference value of the input image that needs to be changed. 
Inspired by~\cite{squid}, we incorporate the use of a random binary mask during training phase to gate shortcut features, thereby enhancing feature aggregation.
For each input image \textit{x}, the final generated image $x^{'}$ following Eq.~\eqref{shortcut} and Eq.\eqref{G}.
\begin{figure}[h]
    \centering
    \includegraphics[width=1.0\textwidth]{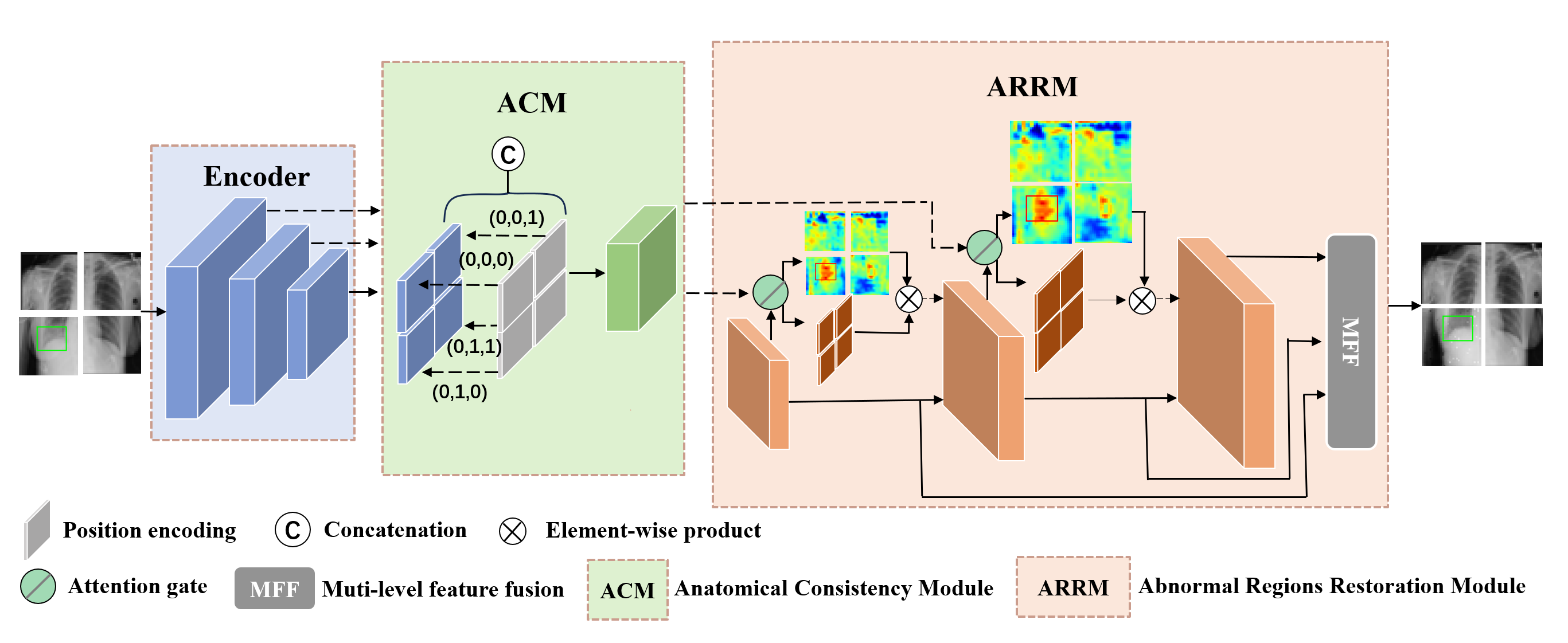}
    \caption{Generator of our proposed SAGAN. The main idea of the generator is to take advantage of the anatomical consistency of medical images and assign higher weights to the attention of anomaly features to refine the restoration of anomaly regions. Generator utilizes positional conditional encoding of patches to extract positional information and uses attention gate to adaptively learn the attention map of regions of interest to accurately control the restoration of anomaly regions.}
    \label{fig:main_pipeline}
\end{figure} 
 
\begin{equation}
\label{shortcut}
    G(x) = I(x)\cdot \delta + (1 - \delta)  \cdot x,
\end{equation}
\begin{equation}
    \label{G}
    x^{'} = tanh(G(x) + x),
\end{equation}
where \textit{G} indicates generator function, $\delta\sim Bernoulli(p)$ is a binary variable with $p$ the gating probability. $tanh$ represents the $tanh$ activation function. Normal images $x_{n}$, pseudo anomaly images $x_{p}$ and unlabeled images $x_{u}$ are passed through the generator to get the corresponding healthy images $x_{n}^{'} $,$x_{p}^{'}$ and $x_{u}^{'}$. We train SAGAN to ensure $x_{u}^{'}$ are close to reality with the supervision of reconstruction of $x_{n}$ and restoration of $x_{p}$. We achieve the training objective by a identity loss, a reconstruction loss and an adversarial loss following Eq.~\eqref{loss_id}, Eq.~\eqref{loss_rec} and Eq.~\eqref{loss_adv_G}. 
\begin{equation}
    \label{loss_id}
    L^{G}_{id} = \mathbb{E}_{x_{1} \in x_{n} } \left \|x_{1}^{'}-x_{1}  \right \| _{1},
\end{equation}
\begin{equation}
    \label{loss_rec}
    L^{G}_{rec} = \mathbb{E}_{x_{2} \in x_{p},x_{1}\in x_{n}} \left \|x_{2}^{'}-x_{1}, \right \| _{2},
\end{equation}
\begin{equation}
    \label{loss_adv_G}
    L^{G}_{adv} = -\mathbb{E}_{x_{3} \in x_{u}} \left [  D(x^{'}_{3})\right ], 
\end{equation}
\begin{equation}
    L^{G} = L^{G}_{adv} + \lambda_{id} L^{G}_{id} + \lambda_{rec} L^{G}_{rec},
\end{equation}
Where $x_{2}$ is the corresponding pseudo anomaly image generated from $x_{1}$. $\lambda_{id}$ and $\lambda_{rec}$ represent the weight of reconstruction loss and identify loss. \textit{D} indicates the discriminator function to distinguish whether the generated image is real or fake. The discriminator classify normal images $x_{n}^{'}$ as real, as well as on healthy images $x_{u}^{'}$ generated from unlabeled images $x_{u}$ as fake. The training objective is achieved using an adversarial loss with a gradient penalty following Eq.~\eqref{loss_adv_D}.
\begin{equation}
    \label{loss_adv_D}
    L^{D}_{adv} = \mathbb{E}_{x_{3} \in x_{u} } \left [ D(x^{'}_{3})\right ] - \mathbb{E}_{x_{1} \in x_{n} } \left [  D(x_{1}) \right ] + \lambda_{gp} \mathbb{E}_{\hat{x}}\left [\left ( \left \| \nabla{\hat{x}}D(\hat{x} ) \right \|_{2}-1  \right )^2 \right ],
\end{equation}
where $\hat{x}$ is a random weighted average of a batch of normal and generated healthy images, and  $\lambda_{gp}$ is the weight of gradient penalty to enhance training stability. In the above process, merely using a simple generator leads to inaccurate restoration. Accordingly, we design a spatial-aware generator consisting of anatomical consistency module (ACM) and abnormal region restoration module (ARRM) to highlight abnormal regions and restore these regions to normal structures at the same position based on spatial information.

\subsection{Positional condition encoding for anatomical consistency}
To fully leverage the anatomical consistency of medical images, we divide the input image into non-overlapping patches of size $N\times N$ and feed them into the generator's encoder for feature extraction. To incorporate positional information and ensure anatomical consistency, we introduce an anatomical consistency module (ACM) adding positional condition encoding for each patch based on its position. Specifically, we initialize a binary code ranging from $0-N^2$. The dimensions of the binary code is determined by $\lceil log_{2}{(N\times N) + 1} \rceil$. Taking $N=2$ as an example, the position of the first patch added is encoded as $[0,0,0]$. Then, the position encoding of patches are extended to corresponding feature. Both extracted features and skip-connections need to add position encoding. Subsequently, given an input image \textit{x}, we refer to the resulting feature vectors as position encoding features $f^{pe} \in \mathbb{R}^{B\times (C+\lceil log_{2}{(N\times N) + 1} \rceil) \times H \times W}$. The spatially-aware features is transformed through the bottleneck layer into $f^{s} \in \mathbb{R}^{B\times C \times H \times W}$.

\subsection{Abnormal region restoration module}
The most significant challenge in the restoration task is accurately restoring abnormal regions.
We define this process as preserving the normal region, `dropping' the abnormal region and generating a corresponding normal one to replace it.
We introduce attention gate to each level of the decoder to achieve this objective. The most important part of attention gate is the attention coefficient, $\alpha _{i} \in  \left [ 0,1 \right ]$, which denotes the scalar attention value for each pixel to identify salient feature and to highlight the abnormal activation. Due to information extracted from coarse scale is used in gating to disambiguate irrelevant and noisy responses in skip connections~\cite{attentiongate}, we choose low-level skip-connections as gating vector $ g^{l}\in \mathbb{R}^{B\times C^{l} \times H^{l} \times W^{l}} $, where $l$ is the number of skip-connection used in training, $C$, $H$ and $W$ denote the channel, height and width of $l^{th}$ skip-connection. Low-level skip-connections highlight anomaly features by increasing the value of the attention coefficient of the anomaly region and decreasing the value of attention coefficient in normal regions. The calculation process of attention gate can be
 expressed as:
 \begin{equation}
     \alpha = S(C_{3}(R(C_{1}(f^{l})+C_{2}(g^{l})))),
 \end{equation}
 \begin{equation}
     g^{l} = \alpha \times g^{l},
 \end{equation}
where $f^{l}$ represent feature of the input of the $l^{th}$ layer of the decoder, $C_{1}$, $C_{2}$ and $C_{3}$ denote convolution layers learning attention coefficients to assign weights to different positions of the parallel input signal. $R$ and $S$ represent ReLU function and sigmoid function. After being filtered through attention gates and decoder, the feature maps are upsampled and fused together using an Muti-level feature fusion module to generate the final restored image with high abnormal response.

\begin{figure}[h]
    \centering
    \includegraphics[width=1.0\textwidth]{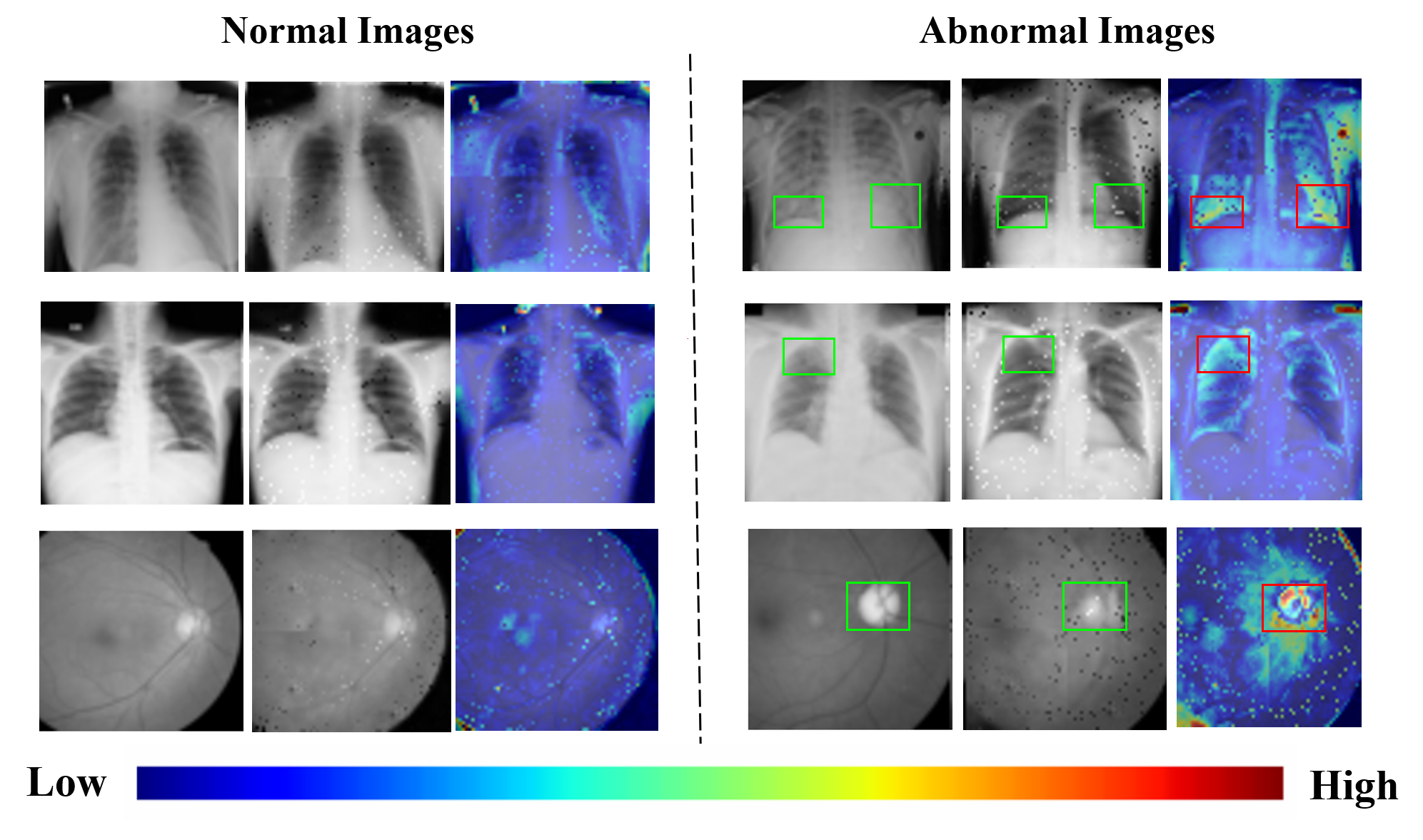}
    \caption{Visualization of heatmaps on medical datasets. The heatmap is derived from the difference maps between the restored image and the original image. Images from top to bottom are from RSNA, VinDr-CXR and LAG datasets, respectively. Original images, restored images, and heatmap visualizations are arranged from left to right. Green boxes represent abnormal regions and red boxes indicate the corresponding abnormal regions on heatmaps.}
    \label{fig:vis}
\end{figure} 

\section{Experiments}

\subsection{Datasets and implementation details}

\textbf{Datasets.} We evaluate our method on two CXR datasets and one retinal fundus image dataset: 1) RSNA Pneumonia Detection Challenge dataset\footnote{\url{https://www.kaggle.com/c/rsna-pneumonia-detection-challenge}}; 2) VinBigData Chest X-ray Abnormalities Detection Challenge dataset (VinDr-CXR)\footnote{\url{https://www.kaggle.com/c/vinbigdata-chest-xray-abnormalities-detection}}; 3) Large-scale Attention-based Glaucoma (LAG) dataset~\cite{li2019attention}. We show a summary of each dataset’s repartitions in Table ~\ref{tab:dataset}. We utilize both normal images $T_{n}$ and unlabeled images $T_{u}$ for model training.

\begin{table}[t]
  \centering
  \caption{Summary of dataset repartitions. Unlabeled image set $T_{u}$ is constructed from the images presented in parentheses without any annotation.}
  \label{tab:dataset}
  \setlength{\tabcolsep}{1mm}
   \resizebox{1\linewidth}{!}{
  \begin{tabular}{llccc}
    \toprule
    \textbf{Dataset} & Normal training set $T_n$ & Unlabeled training set $T_u$ & Testing set $T_{test}$ \\
    \midrule
     
    RSNA & 3851 & 3851 (4000 normal + 5012 abnormal) & 1000 normal + 1000 abnormal \\
     
    VinDr-CXR & 4000 & 4000 (5606 normal + 3394 abnormal) & 1000 normal + 1000 abnormal\\
     
    LAG & 1500 & 1500 (832 normal + 900 abnormal) & 811 normal + 811 abnormal\\
    
    \bottomrule
    
   \end{tabular} 
    }
\end{table}

\begin{table}
    \centering  
    \caption{Performance comparison with SOTA methods. We compare our proposed SAGAN with these state-of-the-art methods which utilize normal images and unlabeled images. We conduct three experiments on each dataset and choose the average value as the final results of the SAGAN. The best results in each category are \textbf{bolded} and the second best results are underlined.} 
    \label{tab::comparsion_sota} 
    \setlength{\tabcolsep}{1mm}
     \resizebox{1.0\linewidth}{!}{
 
	\begin{threeparttable}    
		\begin{tabular}{lcccccccc}  
			\toprule  
			\multirow{2}{*}{\textbf{Method}}& 
            \multirow{2}{*}{\textbf{Taxonomy}}&
			\multicolumn{2}{c}{ \textbf{VinDr-CXR}} & \multicolumn{2}{c}{\textbf{RSNA}} & \multicolumn{2}{c}{\textbf{LAG}} \cr  
			\cmidrule(lr){3-8} 
			&\multicolumn{1}{c}{}&AP(\%)&AUC(\%)&AP(\%)&AUC(\%)&AP(\%)&AUC(\%)\cr  
			\midrule  
            CutPaste~\cite{cutpaste} & Self-sup. &60.0&59.2&61.7&59.8&51.7&48.9\cr
            FPI~\cite{fpi}&Self-sup. &49.4&47.4&53.8&46.6&56.1&52.9\cr
            PII~\cite{pii}&Self-sup.&67.2&66.8&85.4&84.3&63.1&63.1\cr
            NSA~\cite{schluter2022natural}&Self-sup.&64.8&64.4&84.3&84.2&68.0&68.6\cr
            DDAD~\cite{DDAD}&Rec.+ Self-sup&84.3&85.9&91.6&91.3&92.3&93.1\cr
            AMAE~\cite{amae}&Rec.& \underline{84.5}&86.1& \underline{91.7}& \underline{91.4}&-&-\cr
            Brainomaly~\cite{brainomaly}&Restoration.&84.37& \underline{88.54}&87.93&91.01& \underline{94.71}& \underline{95.92}\cr
            \textbf{SAGAN}&Restoration.&\textbf{89.30}&\textbf{91.47}&\textbf{92.60}&\textbf{92.45}&\textbf{96.96}&\textbf{96.98}\cr
			\bottomrule  
		\end{tabular}  
	\end{threeparttable} 
 }
\end{table}

\textbf{Implementation details.} The discriminator network follows PatchGAN~\cite{CGAN,unpaired} architecture and is similar to the ones used in ~\cite{FGAN}. In training stage, we update the weights of the generator once for every two weight updates of the discriminator network. We employ a simple approach to transform a normal image into a corresponding pseudo anomaly image referenced to FPI~\cite{fpi}. We adopt Adam optimizer and set the learning rate and batch size to $5\mathrm{e}^{\scriptstyle -5}$ and 64, respectively. The learning rate has been decayed every 1000 iterations. The maximum epoch is set to 100000 across all settings. For a fair comparison, all the input images are resized to 64$\times$64 pixels. All experiments are implemented using PyTorch and train each models on a single GeForce RTX 2080 Ti GPU. We use the area under the ROC curve (AUC) and average precision (AP) as the evaluation metrics.

\subsection{Main results}
In Table~\ref{tab::comparsion_sota}, we compare our proposed method with a wide range of state-of-the-art (SOTA) methods, including self-supervised synthetic anomaly. We use the unified implementation provided by NSA~\cite{schluter2022natural} for CutPaste, FPI and PII. All other methods used in the experiments are implemented using their official codes. On the RSNA dataset, SAGAN outperforms the previous SOTA method AMAE~\cite{amae} and the baseline method Brainomaly~\cite{brainomaly} by achieving gains of $1.05\%$, $1.00\%$ and $4.67\%$, $1.44\%$ in AP, AUC. On the VinDr-CXR dataset, SAGAN surpass the AMAE by $4.80\%$, $5.37\%$ in AP, AUC, respectively, and outperforms Brainomaly by $4.93\%$, $2.93\%$. On the LAG dataset, SAGAN outperforms the previous SOTA method DDAD~\cite{DDAD} and the baseline method by $4.66\%$, $3.88\%$ and $2.25\%$, $1.06\%$ in AP, AUC. 
We show qualitative results from RSNA dataset, VinDr-CXR dataset and LAG datasets in Fig.~\ref{fig:vis}. Our SAGAN maintains the content of normal images unchanged, while restoring abnormal regions in abnormal images.

\begin{table}[t]
\begin{minipage}[t]{0.55\linewidth}
\captionsetup{labelfont=bf}
\centering
\caption{Ablation study on VinDr-CXR dataset~\cite{li2019attention} for each component of the SAGAN. ACM and ARRM represent Anatomical Consistency Module and Abnormal Regions Restoration Module, respectively.}
\label{tab:ablation}
\begin{tabular}{c|cc}
\toprule
\textbf{Method} & \textbf{AP (\%)$\uparrow$} & \textbf{AUC (\%)$\uparrow$} \\
\midrule
Baseline~\cite{brainomaly} & 84.37 & 88.54 \\
+ ACM & 86.07 & 89.89 \\
+ ARRM & 87.57 & 90.37 \\
+ ACM and ARRM & \textbf{89.30} & \textbf{91.47} \\
\bottomrule
\end{tabular}
\end{minipage}
\hfill
\begin{minipage}[t]{0.4\linewidth}
\renewcommand{\arraystretch}{0.84}
\centering
\caption{Ablation study on RSNA dataset with a varying anomaly ratio (AR) of $T_{u}$. AR = $0_{u}$ denotes experiments trained without unlabeled data.}
\label{tab:ar}
\begin{tabular}{c|cc}
\toprule
\textbf{AR} & \textbf{AP (\%)$\uparrow$} & \textbf{AUC (\%)$\uparrow$} \\
\midrule
$0_{u}$ & 65.87 & 70.54 \\
0 & 75.99 & 77.75 \\
0.2 & 89.65 & 89.23 \\
0.4 & 92.14 & 89.89 \\
0.6 & 92.60 & 92.45 \\
0.8 & 92.64 & 92.42 \\
\bottomrule
\end{tabular}
\end{minipage}
\end{table}

\subsection{Ablation studies}
\textbf{Quantitative analysis of spatial-aware attention.}To understand the effectiveness of each components of SAGAN. We conduct ablation experiments on the VinDr-CXR dataset presented in Table ~\ref{tab:ablation}. It can be clearly seen that the baseline method Brainomaly~\cite{brainomaly} achieve $84.37\%$ AP. Adding the proposed anatomical consistency module, the AP increases by $1.70\%$. On the other, abnormal regions restoration module improved AP by $3.20\%$ suggesting that this module effectively highlights anomaly features. Furthermore,with the addition of these two modules, SAGAN achieved a $4.93\%$ improvement in AP. Accordingly, the results of these experiments demonstrate the effectiveness of each components of SAGAN.


\textbf{Quantitative analysis of anomaly ratio of unlabeled images.} In real-world clinical settings, the anomaly ratio (AR) within vast amounts of unlabeled images is unknown. To simulate the situation, we conduct ablation experiments on the RSNA dataset, varying the AR of $T_{u}$ from $0\%$ to $100\%$, as detailed in Table~\ref{tab:ablation}. In an extreme  experiment, we simulate the absence of unlabeled images by treating pseudo anomaly images as fake in the discriminator. Remarkably, even with an AR of $0\%$, the SAGAN consistently outperforms this scenario, showcasing its ability to enhance restoration performance across all AR situations. Notably, even when the AR of unlabeled data is $20\%$, the SAGAN substantially improves performance, highlighting its capacity to leverage unlabeled images containing anomalies for significant clinical advancements. To ensure fairness in comparison, all experiments are conducted under AR = $60\%$.

\section{Conclusion}
In this paper, we propose the Spatial-aware Attention Generative Adversarial Network for anomaly detection (SAGAN), which fully utilizes anomaly spatial features in unlabeled images to efficiently restore anomaly regions. Experiments on three benchmarks demonstrate that SAGAN achieves state-of-the-art performance. As for the limitation, SAGAN is not able to localize anomalies at the pixel level precisely and SAGAN still requires a generous number of normal annotations in training stage. In the future work, we will extend our SAGAN method to achieve accurate anomaly localization and explore the potential of anomaly detection without any training annotations.

\bibliographystyle{splncs04}
\bibliography{reference}

\end{document}